\documentstyle[12pt]{article}
\newlength{\defaultparindent}
\begin{document}
\subsection*{
Exact diagonalization of the S = 1/2 Heisenberg
antiferromagnet on finite bcc lattices to estimate properties
on the infinite lattice }

\noindent 
D.D.~Betts$^{1\dagger}$, J.~Schulenburg$^2$, 
G.E.~Stewart$^3$, J.~Richter$^{2\ast}$ and J.S. Flynn$^1$

\vspace{5mm}\noindent {\small
$^1$ Department of Physics, Dalhousie University, Halifax, NS, Canada B3H 3J5
\\ $^2$ Institut f\"ur Theoretische Physik, Universit\"at Magdeburg,
  PO Box 4120, D-39016, Magdeburg, Germany
\\ $^3$ Pulp and Paper Centre, University of British Columbia, 2385
  East Mall, Vancouver, BC, Canada V6T 1Z4 }

\vspace{5mm}\noindent
Received . . . May 1998

\vspace{5mm}\noindent
Short Title: Diagonalize Heisenberg model on bcc lattices

\vspace{5mm} \noindent
{\bf Abstract.} This paper is the third in a sequence of papers developing
and applying the finite lattice method for estimation of the
zero temperature properties of quantum spin models on infinite
cubic lattices.  Here we generate finite bipartite body-centred
cubic lattices of $16 \le N \le 32$ vertices.  Our
geometrically distinct finite lattices are defined by vectors
in upper triangular lattice form.  We have found that sets of
two to six geometrically distinct finite bcc lattices are topologically
identical, and that we thus need only one lattice of each set
for our method of estimation.  We have studied the spin one half
Heisenberg antiferromagnet by diagonalizing its Hamiltonian on
each of the finite lattices and hence computing its ground state
properties.  By extrapolation of these data we obtain estimates
of the T = 0 properties on the infinite bcc lattice.  Our estimate
of the T = 0 energy agrees to five parts in ten thousand with
third order spin wave and series expansion method estimates,
while our estimate of the staggered magnetization agrees with
the spin wave estimate to within a quarter of one percent.

\vfill

\begin{flushleft}
 $^\dagger$ E-mail address: dbetts@is.dal.ca\\
 $^\ast$ E-mail address:  johannes.richter@physik.uni-magdeburg.de
\end{flushleft}

\newpage
\subsubsection*{1. Introduction}

The physics of quantum spin systems on lattices has been much
studied and remains of great interest.  It is a major part of
condensed matter physics.  In particular at zero temperature
questions such as what the energy per vertex is, whether long
range order exists, if so its nature, the possibility of quantum
phase transitions, the spatial dependence of spin-spin correlations,
etc. are being studied.  Several different methods of calculating
the properties of quantum spin systems such as the Heisenberg
antiferromagnet, the XY ferromagnet and the t-J model are being
developed and used.  The usefulness of a method of calculation
of properties depends greatly on the dimension, d, of the lattice
and the temperature, T.

At T = 0 and for d = 2 or 3 series expansions starting from the
Ising model limit and spin wave methods have proved to be particularly
useful [1].  The quantum Monte Carlo method would be very useful
too, but we are unaware of its application to quantum spin models
on cubic lattices.

In 1964 Bonner and Fisher [2] introduced what we could now call
the method of exact diagonalization on finite lattices in one
dimension.  Two decades ago Oitmaa and Betts introduced the finite
lattice method in two dimensions [3,4].  It became popular after
high Tc superconductivity was discovered.  For further information
recent review articles are recommended [5,6].  However, it seems
that until recently (except for one short exploratory paper [7])
no one extended the method of exact diagonalization on finite
lattices to three dimensions.

In 1997 Betts and Stewart published their paper on estimation
of zero-temperature properties of quantum spin systems on the
simple cubic lattice via exact diagonalization on finite simple
cubic lattices [8].  This method was soon extended to finite
face centred cubic lattices for the estimation of the T = 0 properties
of the spin one half XY ferromagnet on the infinite fcc lattice
using finite fcc lattices of $N \le 25$ vertices [9].  The
finite lattice method estimates of the energy and magnetization
per vertex of this model agreed to within a fraction of a percent
with the estimates by spin wave, series expansion and variational
methods.

We learned from Lyness et al [10] about using a triple of vectors
in upper triangular lattice form (utlf) to define finite lattices
in three dimensions.  Generating finite lattices in this way
ensures that each lattice is geometrically distinct.  Some of
us used the utlf method first to define finite fcc lattices [9].
 Since then we have learned that some lattices that are geometrically
distinct are topologically identical.  Hence, as far as we have
investigated, each physical property of each quantum spin model
with nearest neighbour interactions has the same numerical value
on all topologically identical though geometrically distinct
lattices.

In Sec. 3 we describe how to classify finite lattices topologically.
 First we define a topological neighbourhood matrix and calculate
one for each finite bcc lattice.  Any two geometrically distinct
finite lattices whose neighbourhood matrices are equivalent are
topologically identical lattices.  We find such equivalence in
two separate ways.  Our topological sorting code is very easy
to calculate but is useful only for bipartite finite lattices
of N $<$ 32.  The little known Smith normal form matrix can work
for large as well as small lattices but is more complicated to
calculate.

Section 4 describes the computation of the ground state eigenvalue
and eigenvector of the spin one-half Heisenberg Hamiltonian on
the finite bcc lattices.  Ground-state spin-spin correlations
are computed for each lattice using its eigenvector.  Thence
the staggered magnetization is calculated.  Statistical analyses
fit formulas, determined by spin wave theory [1], in N$^{-1/3}$ to
the data in turn for energy, magnetization and correlations and
thus provide finite lattice method estimates of the T = 0 properties
on the infinite bcc lattice.  Since our method includes the precise
calculation of the ground state of the eigenvector, it has the
advantage of the ready calculation of any property of the quantum
spin model based on its ground state eigenvector.

Finally we compare our finite results with the recent series
expansion, third order spin wave [1] and variational [11] estimates.
 Our estimate of the energy per vertex agrees with the spin wave
and series expansion estimates very closely.  Our estimate of
the staggered magnetization agrees to within a quarter of one
percent with the estimates from both of the above methods.  The
variational estimate of energy is one percent below the other
three estimates, and the variational estimate of staggered magnetization
is three percent below the other estimates.

\subsubsection*{2. Generation of Bipartite Finite Body-Centred Cubic Lattices}

The unbounded bcc lattice can be defined by any three of the
primitive vectors ${\bf a}_1 = (1,1,1)$, ${\bf a}_2 = (1,1,-1)$,
  ${\bf a}_3 = (1,-1,1)$ 
and  ${\bf a}_4 = (-1,1,1)$.  The unbounded
bcc lattice can be filled by identical parallelepipeds each defined
by three edge vectors,

\[ {\bf l}_\alpha = \sum_{\beta=1}^3 n_{\alpha\beta}{\bf a}_\beta
   \hspace{5cm} (2.1) \]
where $n_{\alpha\beta}$ are integers.

A finite bcc lattice can be derived from any one of the parallelepipeds
described above by being subjected to periodic boundary conditions.
 That is, each of the three pairs of opposite faces are identified.

A finite bcc lattice is bipartite if each of the nine coefficients,
$n_{\alpha\beta}$, is an even integer.  The resulting bipartite finite bcc
lattice consists of two identical finite simple cubic lattices.
 Any of several different parallelepipeds can generate the same
finite lattice.  Another way of regarding a finite lattice is
to consider an unbounded lattice, bcc, fcc or whatever, as being
composed of N identical sublattices.  Then on each of the N sublattices
all vertices are considered to be one vertex.

In Figure 1 we provide an example of a bipartite finite bcc lattice.
 The parallelepiped which by replication fills the unbounded
bcc lattice is defined by edge vectors

\[ {\bf L}_1 = ( 0,-2, 4), \hspace{3mm}
   {\bf L}_2 = ( 2, 4, 0), \hspace{3mm}
   {\bf L}_3 = (-2, 4, 2)  \hspace{2cm} (2.2) \]
The volume of this parallelepiped is 72, which means that it
contains N = 18 vertices.  The vertices on one simple cubic sublattice
are represented by black circles and those on the other sublattice
by white squares.  Of course, each of the eight corner vertices
is shared by eight parallelepipeds, and each of the two face-centred
``white vertices'', (-1, 1, 3) and (1, 5, 3), is shared between
two parallelepipeds.

An N = 18 bipartite finite bcc lattice is formed by the application
of periodic boundary conditions in all three directions, that
is, by identifying each pair of opposite faces of the parallelepiped.
 The resulting nine distinct vertices on one sublattice are labelled
A, B, C, ... I and the nine vertices on the other sublattice
are labelled a, b, c, ... i.  We have drawn bonds only between
vertex ( 0, 4, 2) and its eight nearest neighbours. ( Drawing
all the bonds would make Figure 1 appear too cluttered. )  Note
that two of these bonds each have two pieces within the parallelepiped,
but because of periodic boundary conditions each of these bonds
is continuous in the finite lattice.

Since publishing the first paper on exact diagonalization on
finite cubic lattices [8] we learned about describing the generating
of finite lattices by parallelepipeds in upper triangular lattice
form (utlf).  An $n \times n$ matrix, B, is in upper triangular lattice
form [10] if and only if the integral matrix elements, $b_{ij}$ ,
satisfy the following criteria:

\begin{tabbing}
xxxxx \= xxxxxxxxxxxxxxxx             \= xxxxxxxxxxxxxxxxxxxxxxx \= \kill
      \> $b_{ii} \ge 1$               \> i = 1,2, ...n \\
      \> $b_{ij} = 0 $                \> $ 1 \le j < i \le n $ \> (2.3) \\
      \> $b_{ij} \in  [0, b_{jj}) $   \> $ 1 \le i < j \le n $ \\
\end{tabbing}

For utlf parallelepipeds on the bcc lattice the first edge vector,
${\bf l}_1$, is in the positive octant of  the lattice space but not in
the yz plane; the second edge vector , ${\bf l}_2$, lies in the positive
quadrant of the yz plane but not along the z axis; the third
edge vector, ${\bf l}_3$ , lies on the positive z axis.  These three edge
vectors form a $3 \times 3$ utlf matrix, B.  By using only generating,
or defining, vectors in utlf form each finite lattice is described
only once, so we have no need to search for duplicates.  Further
discussion on utlf matrices is contained in Stewart et al [9]
and Lyness et al [10].

Here we describe our best criteria for the initial selection
of finite bipartite bcc lattices. First, each vertex must have
eight distinct nearest neighbours for a model such as the usual
Heisenberg antiferromagnet. Thus all finite bipartite bcc lattices
must have at least sixteen vertices. At the other limit one need
not generate bcc lattices of more vertices than the computer
can deal with in a reasonable length of time; in our present
case N $<$ 33. A useful lattice of N $<$ 31 should have no pair
of vertices farther apart than geometrically seventh neighbours
or equivalently topologically third neighbours (as described
below). For 31 $<$ N $<$ 41 lattices topologically fourth neighbours
should be included. Such criteria for the bcc and other lattices
can be applied in a simple computer program. With the computing
facilities available to us, we could complete the exact diagonalization
of the antiferromagnetic S = 1/2 Heisenberg Hamiltonian on finite
bcc lattices of more than 32 vertices, but it would have taken
much more time.  However, we are pleased to have computed the
ground state eigenvector of sixteen topologically distinct bcc
lattices of 32 vertices.

Table 1 gives, for 16 $\le$ N $\le$ 32, the number of
finite bcc lattices that are geometrically distinct, the number
of the geometrically distinct lattices that are bipartite, the
number of those that are also topologically distinct and finally,
the number of those remaining which are also statistically useful
in our method.

\begin{table}
Table 1. Numbers of finite bcc lattices of even N vertices in
which each lattice has eight distinct nearest neighbour vertices.
 The number headings are:  $n_g$ - geometrically distinct lattices;
$n_b$ - bipartite lattices; $n_t$  - topologically distinct bipartite
lattices; $n_u$  - statistically useful lattices.

\begin{center}\begin{tabular}{rp{1cm}p{1cm}p{1cm}p{1cm}p{1cm}} \hline\hline
 N && $n_g$ & $n_b$ & $n_t$ & $n_u$ \\ \hline
 16 &   & 12    & 5  & 1  & 1 \\
 18 &   & 15    & 1  & 1  & 1 \\
 20 &   & 22    & 4  & 2  & 2 \\
 22 &   & 16    & 1  & 1  & 1 \\
 24 &   & 52    & 15 & 6  & 5 \\
 26 &   & 21    & 2  & 2  & 2 \\
 28 &   & 44    & 9  & 6  & 4 \\
 30 &   & 52    & 7  & 7  & 5 \\
 32 &   & 57    & 20 & 16 & 10 \\ \hline
Totals && 291   & 64 & 42 & 31 \\ \hline
\end{tabular}\end{center}
\end{table}

All topologically distinct bipartite body-centred cubic lattices
of 16 $\le$  N $\le$ 32 vertices are described in Table 2 below. 
Each lattice is described by a label N$\alpha$.  The ``best'' lattice for each N
is labelled A, the second best B, etc..  The statistical analysis
of the energy data for the Heisenberg antiferromagnet described
in Section 4 determines the order of goodness of the finite lattices.
 The finite lattices are defined by a set of three vectors, ${\bf l}_i$,
in upper triangular lattice form.  One can consider the three
vectors as defining the edge of a parallelepiped, and then that
the three pairs of opposite faces are identified, thus defining
the finite lattice.  The same finite lattice can then be defined
by a different ``compact'' parallelepiped whose compact edge
vectors, ${\bf L}_j$, are as short as possible

\[ {\bf L}_j=\sum_{i=1}^3 m_{ji} {\bf l}_i \hspace{5cm}
                                        (2.4) \]
with each coefficient, $m_{ji}$, being an integer, positive, negative,
or zero.  The converse linear relation also requires integral
coefficients.

\begin{table} {\small
Table 2.  Defining vectors of topologically distinct bipartite
bcc lattices of $16 \le N \le 32$ vertices. 
($^\ast$ These finite lattices were statistically found to be outriders.)
\begin{center}
\begin{tabular}{l l ccc p{3mm} ccc} \hline\hline
 &  N$\alpha$ & \multicolumn{3}{c}{ utlf vectors }  
 &            & \multicolumn{3}{c}{ compact vectors } \\ \cline{3-5} \cline{7-9}
 &            & ${\bf l}_1$ & ${\bf l}_2$ & ${\bf l}_3$  
 &            & ${\bf L}_1$ & ${\bf L}_2$ & ${\bf L}_3$ \\ \hline
 & 16A & 	(4,0,0) & (0,4,0) & (0,0,4) && (4,0,0) & (0,4,0) & (0,0,4) \\ \hline

 & 18A & 	(2,0,8) & (0,2,4) & (0,0,18)&& (2,-2,4)& (0,2,4) & (4,0,-2)\\ \hline
 
 & 20A & 	(2,0,10)& (0,2,6) & (0,0,20)&& (2,-2,4)& (-2,4,2)& (4,0,0) \\
 & 20B & 	(2,0,8) & (0,2,4) & (0,0,20)&& (2,-2,4)& (0,2,4) & (4,0,-4)\\ \hline

 & 22A & 	(2,0,8) & (0,2,4) & (0,0,22)&& (2,-2,4)& (0,2,4) & (4,2,-2)\\ \hline

 & 24A & 	(2,0,8) & (0,2,4) & (0,0,24)&& (2,-2,4)& (0,2,4) & (4,4,0) \\
 & 24B & 	(2,0,4) & (0,4,4) & (0,0,12)&& (2,0,4) & (0,4,4) & (4,4,0) \\
 & 24C & 	(2,0,10)& (0,2,6) & (0,0,24)&& (2,-2,4)& (-2,4,2)& (4,2,-2)\\
 & 24D & 	(2,0,4) & (0,4,6) & (0,0,12)&& (2,0,4) & (-2,4,2)& (2,4,-2)\\
 & 24E & 	(2,0,12)& (0,2,4) & (0,0,24)&& (2,-4,4)& (0,2,4) & (4,0,0) \\
 & 24F$^\ast$& (2,0,12)& (0,2,6) & (0,0,24)&& (4,0,0) & (2,-4,0)& (2,2,-6)\\ \hline

 & 26A & 	(2,0,10)& (0,2,4) & (0,0,26)&& (2,-4,2)& (0,2,4)& (4,2,-2)\\
 & 26B & 	(2,0,8) & (0,2,4) & (0,0,26)&& (2,-2,4)& (0,2,4)& (4,4,-2)\\ \hline

 & 28A & 	(2,0,12)& (0,2,8) & (0,0,28)&& (2,-2,4)& (-2,4,4)& (2,4,0)\\
 & 28B & 	(2,0,12)& (0,2,4) & (0,0,28)&& (2,-4,4)& (0,2,4) & (4,2,0)\\
 & 28C & 	(2,0,10)& (0,2,4) & (0,0,28)&& (2,-4,2)& (0,2,4) & (4,4,0)\\
 & 28D & 	(2,0,10)& (0,2,6) & (0,0,28)&& (2,-2,4)& (-2,4,2)& (4,2,-2)\\
 & 28E$^\ast$& (2,0,8) & (0,2,4) & (0,0,28)&& (-2,4,0)& (0,2,4) & (4,4,-4)\\
 & 28F$^\ast$& (2,0,14)& (0,2,4) & (0,0,28)&& (4,0,0) & (0,2,4) & (2,-6,2)\\ \hline
	
 & 30A & 	(2,0,12)& (0,2,4) & (0,0,30)&& (2,-4,4)& (0,2,4) & (4,2,-2)\\
 & 30B & 	(2,0,4) & (0,6,0) & (0,0,10)&& (2,0,4) & (4,0,-2)& (0,6,0)\\
 & 30C & 	(2,0,12)& (0,2,6) & (0,0,30)&& (2,-4,0)& (0,2,6) & (4,2,0)\\
 & 30D & 	(2,0,12)& (0,2,8) & (0,0,30)&& (2,-2,4)& (-2,4,4)& (4,2,2)\\
 & 30E & 	(2,0,10)& (0,2,4) & (0,0,30)&& (2,-4,2)& (0,2,4) & (4,4,-2)\\
 & 30F$^\ast$& (2,0,10)& (0,2,6) & (0,0,30)&& (2,-4,0)& (-2,4,2)& (4,2,-4)\\
 & 30G$^\ast$& (2,0,8) & (0,2,4) & (0,0,30)&& (2,-4,0)& (0,2,4) & (6,2,-2)\\ \hline

 & 32A & 	(2,0,14)& (0,2,8)& (0,0,32)&& (2,-2,6)& (-2,4,2) & (2,4,-2)\\
 & 32B & 	(2,2,6) & (0,4,0)& (0,0,16)&& (2,2,6) & (0,4,0)  & (4,4,-4)\\
 & 32C & 	(2,0,4) & (0,4,8)& (0,0,16)&& (2,0,4) & (-2,4,4) & (4,4,0)\\
 & 32D & 	(2,2,4) & (0,8,0)& (0,0,8) && (2,2,4) & (2,-6,4) & (4,4,0)\\
 & 32E & 	(2,0,10)& (0,2,4)& (0,0,32)&& (2,-4,2)& (0,2,4)  & (4,4,-4)\\
 & 32F & 	(2,0,6) & (0,4,8)& (0,0,16)&& (-2,4,2)& (4,-4,4) & (2,4,-2)\\
 & 32G & 	(2,0,12)& (0,2,4)& (0,0,32)&& (2,-4,4)& (0,2,4)  & (4,4,0)\\
 & 32H & 	(2,0,10)& (0,2,6)& (0,0,32)&& (2,-2,4)& (-2,4,2) & (4,4,0)\\
 & 32I & 	(2,0,12)& (0,2,8)& (0,0,32)&& (2,-2,4)& (-2,4,4) & (4,2,0)\\
 & 32J$^\ast$& (4,0,4) & (0,4,4)& (0,0,8) && (4,0,4) & (0,4,4)  & (4,4,0)\\
 & 32K$^\ast$& (2,0,6) & (0,4,4)& (0,0,16)&& (4,4,0) & (0,4,4)  & (2,-4,2)\\
 & 32L$^\ast$& (2,4,4) & (0,8,0)& (0,0,8) && (0,4,4) & (2,4,-4) & (2,4,4)\\
 & 32M$^\ast$& (2,0,14)& (0,2,4)& (0,0,32)&& (4,2,0) & (0,2,4)  & (2,-6,2)\\
 & 32N$^\ast$& (2,0,4) & (0,4,4)& (0,0,16)&& (2,0,4) & (0,4,4)  & (4,4,-4)\\
 & 32P$^\ast$& (2,0,8) & (0,2,4)& (0,0,32)&& (2,-2,4)& (0,2,4)  & (2,-6,4)\\
 & 32Q$^\ast$& (2,0,16)& (0,2,4)& (0,0,32)&& (4,0,0) & (0,2,4)  & (2,-6,4)\\ \hline
\end{tabular}
\end{center}
}
\end{table}

\subsubsection*{3.  Topologically Distinct Finite Lattices}

After publication of the first article on using finite three
dimensional (simple cubic) lattices to estimate the zero temperature
properties of quantum spin systems [8] we became puzzled over
the computed properties of some of the finite lattices.   An
examination of Table 1 in that publication shows that three apparently
geometrically distinct N = 16 bipartite simple cubic lattices
have the same computed ground-state energy and the same long
range order for the Heisenberg antiferromagnet and for the XY
ferromagnet.  This phenomenon also occurs for a pair of N = 18
bipartite simple cubic lattices and a triple of N = 20 bipartite
s.c. lattices.

In our second paper [9], developing the finite lattice method
on the face-centred cubic lattice for the XY ferromagnet, no
such phenomenon was found.  However, there are no bipartite fcc
lattices.  When we turned to the Heisenberg antiferromagnet on
the bcc lattice the same phenomenon was manifest.  We wondered
whether such geometrically distinct finite lattices, simple cubic
or bcc, could be topologically identical.

Our first step in determining whether two geometrically distinct
finite lattices of N vertices are also topologically distinct
is to construct a {\it topological neighbourhood matrix} for each
finite lattice.  We label the vertices a,b,c, ... on one sublattice
and A,B,C, ... on the other.  Suppose  an imaginary microscopic
frog can hop from one vertex to only one of its nearest neighbour
vertices.  If the frog has to make a minimum of h hops to get
from vertex a to vertex b,  then these two vertices are topologically
h'th neighbours to one another.  (Of course,  each vertex is
a zeroth neighbour to itself.)  Both the columns and rows of
the neighbourhood matrix are labelled alphabetically.  Thus matrix
element ab equals matrix element ba, and this element is the
positive integer h.

Two finite lattices with neighbourhood matrices U and V are defined
to be topologically identical if and only if there exists a permutation
matrix P such that UP = PV.

In particular, we are interested here in determining whether
two geometrically distinct finite bipartite bcc lattices are
topologically identical or distinct.  If the initial vertex is
at the origin, then the eight nearest neighbours  would have
coordinates $(\pm 1,\pm 1,\pm 1)$.  The twenty-six topologically second
neighbours to the origin would have coordinates of one of the
three types (2,0,0), (2,2,0), or (2,2,2), and the fifty-six topologically
third neighbours would have coordinates of the types (3,1,1),
(3,3,1), or (3,3,3).  None of our finite bcc lattices of N $<$
32 has any pair of topologically fourth neighbour vertices.
Thus, for any bipartite finite bcc lattice of    N $<$ 32, every
pair of vertices on the same sublattice are topologically second
neighbours.  One vertex on one sublattice and one on the other
sublattice are topologically either nearest or third neighbours.

As an example, we consider the four geometrically distinct N
= 20 bipartite bcc lattices.  We label the vertices on one sublattice
A,B, ..., J and those on the other sublattice a,b, ...,j.
 The rows and the columns of the 20 $\times$ 20 neighbourhood matrices
are labelled in the above order.  Since every off-diagonal element
in the upper left quadrant and the lower right quadrant of the
neigbourhood matrix of each of the four lattices is 2, it is
sufficient to consider only the distinguishable lower left (or
upper right) quadrants.  Three of the submatrices are displayed
in Table 3.

\begin{table}
Table 3.  Lower left quadrant of the topological neighbourhood
matrix of three N=20 bipartite bcc lattices.
\begin{center}\begin{tabular}{llccc} \hline\hline
&   &  lattice 20.1              &  lattice 20.2              &  lattice 20.3            \\
&   &  ${\bf l}_1$ = (2,0, 8)    &  ${\bf l}_1$ = (2,0,10)    &  ${\bf l}_1$ = (2,0,10)  \\
&   &  ${\bf l}_2$ = (0,2, 4)    &  ${\bf l}_2$ = (0,2, 4)    &  ${\bf l}_2$ = (0,2, 6)  \\
&   &  ${\bf l}_3$ = (0,0,20)    &  ${\bf l}_3$ = (0,0,20)    &  ${\bf l}_3$ = (0,0,20)  \\ \hline
&   &{\tt  A B C D E F G H I J }&{\tt  A B C D E F G H I J  }&{\tt  A B C D E F G H I J}\\
& a &{\tt  1 1 1 1 3 3 1 1 1 1 }&{\tt  3 1 1 1 1 3 1 1 1 1  }&{\tt  1 1 3 1 1 1 1 3 1 1}\\
& b &{\tt  1 1 1 1 1 3 3 1 1 1 }&{\tt  1 3 1 1 1 1 3 1 1 1  }&{\tt  1 1 1 3 1 1 1 1 3 1}\\
& c &{\tt  1 1 1 1 1 1 3 3 1 1 }&{\tt  1 1 3 1 1 1 1 3 1 1  }&{\tt  1 1 1 1 3 1 1 1 1 3}\\
& d &{\tt  1 1 1 1 1 1 1 3 3 1 }&{\tt  1 1 1 3 1 1 1 1 3 1  }&{\tt  3 1 1 1 1 3 1 1 1 1}\\
& e &{\tt  1 1 1 1 1 1 1 1 3 3 }&{\tt  1 1 1 1 3 1 1 1 1 3  }&{\tt  1 3 1 1 1 1 3 1 1 1}\\
& f &{\tt  3 1 1 1 1 1 1 1 1 3 }&{\tt  3 1 1 1 1 3 1 1 1 1  }&{\tt  1 1 3 1 1 1 1 3 1 1}\\
& g &{\tt  3 3 1 1 1 1 1 1 1 1 }&{\tt  1 3 1 1 1 1 3 1 1 1  }&{\tt  1 1 1 3 1 1 1 1 3 1}\\
& h &{\tt  1 3 3 1 1 1 1 1 1 1 }&{\tt  1 1 3 1 1 1 1 3 1 1  }&{\tt  1 1 1 1 3 1 1 1 1 3}\\
& i &{\tt  1 1 3 3 1 1 1 1 1 1 }&{\tt  1 1 1 3 1 1 1 1 3 1  }&{\tt  3 1 1 1 1 3 1 1 1 1}\\
& j &{\tt  1 1 1 3 3 1 1 1 1 1 }&{\tt  1 1 1 1 3 1 1 1 1 3  }&{\tt  1 3 1 1 1 1 3 1 1 1}\\ \hline
\end{tabular}\end{center} 
\end{table}

A visual inspection of such small submatrices as in Table 3 will
reveal which pairs are equivalent and which are distinct.  Notice
that the middle matrix representing lattice 20.2 consists of
 ``quartets'' of vertices.  For example, vertices A and F on
one sublattice have in common the same two third neighbour vertices,
a and f, on the other sublattice.  Each of the vertices belongs
to one and only one such quartet.  The matrix on the right demonstrates
the same quartet structure for lattice 20.3.  Obviously these
two lattices are equivalent under permutation, and so the two
lattices they represent are topologically identical.  However,
the matrix on the left shows that any two vertices on one sublattice
of lattice 20.1 have one and only one third neighbour in common
on the other sublattice.  Finite lattice 20.1 is topologically
distinct from lattices 20.2 and 20.3. The computed ground state
properties of the Heisenberg antiferromagnet, i.e. the energy,
staggered magnetization, and spin-spin correlations, each of
them having different values on lattice 20.1 from the identical
values on lattices 20.2, 20.3, and 20.4, confirm the topological
classification derived by inspecting the neighbourhood matrices.
 (Lattice 20.4 is defined by the vectors ${\bf l}_1$ = (2,0,4), 
 ${\bf l}_2$ = (0,4,0), and ${\bf l}_3$ = (0,0,10).)

However, for larger N, simple inspection of the neighbourhood
matrices is inadequate to sort the finite bipartite lattices
topologically.  Thus we define a nine-digit {\it sorting code} based
on the lower left quadrant of each neighbourhood matrix.  For
each finite lattice choose one of the sublattices and one of
the N/2 vertices on it, say A.  Each vertex on the chosen sublattice,
A included, will have zero to eight nearest neighbours on the
other sublattice that are also nearest neighbours to A.  The
$v_i$ vertices on the chosen sublattice will have i nearest neighbours
in common with A. The topological sorting code is defined as

\[ C (\mbox{index}) = v_0, v_1, . . . v_8 \hspace{5cm}
           (3.1) \]
In Table 3 it is easy to see for lattice 20.1 that vertices B
and J have seven nearest neighbours in common with A and the
other seven vertices on the same sublattice have six nearest
neighbours in common with A.  (Of course, A has eight nearest
neighbours in common with A).  Thus the topological sorting code
C(20.1) = 000~000~721.  In lattices 20.2, 20.3, and 20.4 vertices
F and A have eight nearest neighbours in common and each of the
other eight vertices on this sublattice have six nearest neighbour
in common with A; hence code C(20.2) = 000~000~802 = C(20.3)
= C(20.4).  Thus these three lattices are topologically identical,
according to the code.  We have confirmed this identity, and
several other identities, via computed values of the energy,
staggered-magnetization and spin-spin correlations of the Heisenberg
antiferromagnet.  In some cases further confirmation of topological
identity or distinction has been reinforced by the energy, magnetization
and spin-spin correlations of the S = 1/2 XY ferromagnet.

We have used another criterion for the topological classification
of finite lattices, the Smith normal form [12].  For any square
matrix, A, of finite rank, n, there exists just one matrix, F,
also of rank n, in Smith normal form.

\[ F=\left( \begin{array}{cc} D & 0\\ 0 & 0\\ \end{array} \right)
 \hspace{5cm}                     (3.2)     \]
where $D$ is a diagonal matrix of rank $r \le n$.  The elements
on the diagonal, 

\[ d_1 = b_1 = 1,   d_2  = b_1 b_2 , . . . d_r = b_1  b_2 . . . b_r
 \hspace{4cm}      (3.3) \]
The neighbourhood matrices of interest here are in the field
of positive integers so $b_i$ is integral.

All equivalent matrices have the same Smith normal form, but
the Smith matrix is unique to that one set of equivalent matrices.

The definition of the $b_i$ is complicated and so is omitted here.
 However, the definition of $b_i$, the process of calculating a
Smith normal form matrix and the proof of the above properties
of the Smith normal form can be found in several texts on matrices,
e.g. Turnbull and Aitken [13].   Hand calculation of the Smith
normal form of a matrix of only integer elements and a rank as
low as eight would be extremely tedious.  However, programs to
compute the Smith normal form for much larger matrices are readily
available.  We have used the Maple program.  

As an example, here are the diagonal elements of the Smith normal
forms for the lower left quadrant of the neighbourhood matrices
of the four N = 20 bipartite bcc lattices:

\[ F (20.1) = 1,2,2,2,2,2,2,2,14,0 \hspace{8mm}\mbox{and} \]
\[ F (20.2) = F(20.3) = F(20.4) = 1,2,2,2,14,0,0,0,0,0 \]
It turns out that neither of the two methods, {\it by themselves},
are completely successful in finding all of the topologically
identical sets of lattices.  However, combining the two methods
appears to find all of these sets.  This was confirmed by examining
the computed properties of the S = 1/2 Heisenberg antiferromagnet
on bipartite bcc lattices of N $\le$ 30.  Thus we were able
to greatly reduce the computer time to diagonalize this model
on N = 32 lattices by using only one lattice of each topology
as indicated by our sorting codes and Smith normal forms.

\subsubsection*{4.  Computation of the ground state properties of the S = 1/2
Heisenberg antiferromagnet on finite bcc lattices and statistical
estimates of the zero temperature properties of this model on
the infinite bcc lattice}

The Hamiltonian of the spin one half Heisenberg antiferromagnet
in zero field is

\[ H=-J\sum_{\langle i,j\rangle} {\bf S}_i \cdot {\bf S}_j 
 \hspace{5cm} (4.1) \]
where the sum is over nearest neighbour pairs of vertices.  It
was proved by Lieb and Mattis [14] that the ground state of this
model on a bipartite three dimensional lattice has total spin
equal to zero and is nondegenerate.  Later it was proved by Kennedy
et al [15] and Kubo and Kishi [16] that this model has long range
N\`{e}el order in the ground state.

All our finite lattices and thus the corresponding Hamiltonians
are translationally invariant as well as invariant under inversion,
which simplifies the diagonalization of the Hamiltonians.

The diagonalization of the Hamiltonian matrix to obtain the ground
state eigenvalue (the energy) and the ground-state eigenvector
has been done mostly by workstations and a Power Challenge computer
at the University of Magdeburg with some input from Dalhousie's
SP2 computer.  The Lanczos technique used in the diagonalization
is standard [17].  In order to diagonalize the Hamiltonian on
the larger lattices we had to reduce the dimensions of the Hilbert
space by using the translation and point group symmetries of
the Hamiltonian.  The largest Hamiltonian we diagonalized, on
an N = 32 lattice, is of rank 4.7 million approximately.  Due
to the limited precision of the computer, the precision of the
ground state eigenvalue (or energy) of the Hamiltonian on the
larger lattices is 7 or 8 digits.

Using the ground state eigenvector we have computed the ground
state spin-spin correlations, 

$<{\bf S}_i\cdot {\bf S}_j>$,
for all pairs of spins on all geometrically distinct
bipartite bcc lattices of N vertices where 16 $\le$ N $\le$
30, and for N = 32 we have computed the energies and correlations
of all {\it topologically} distinct bipartite lattices only.  

The principal results are displayed in Table 4 for all topologically
distinct bipartite bcc lattices of 16 $\le$ N $\le$ 32.
For each lattice only the average correlations are displayed
for topologically first, second and third neighbour correlations.
 We have omitted fourth neighbour correlations from Table 4 because
only some bcc lattices of N $\ge$ 32 have fourth neighbour pairs
of vertices.  The average of first neighbour correlations is
simply the ground state energy per vertex divided by 4J.

\begin{table} {\small
Table 4. The data below include the staggered magnetization
per vertex, $m^+$, and first, second and third topological neighbour
spin-spin correlations, $\Gamma_1$, $\Gamma_2$ and $\Gamma_3$, 
and the statistical weight,
$w_1$, of $\Gamma_1$ (or the ground state energy), of the S = 1/2 Heisenberg
antiferromagnet on all topologically distinct finite bcc lattices
of $16 \le N \le 32$
\begin{center}\begin{tabular}{ll l l l l l} \hline\hline
& N$\alpha$ &  $m^+$ &  $\Gamma_1$ &   $w_1$ &   $\Gamma_2$ &  $\Gamma_3$ \\ \hline

& 16A&  0.559 0170&  -0.312 500&  0.996&  0.250 000&  ---\\ \hline
 
& 18A&  0.549 6565&  -0.309 415&  0.999&  0.246 137&  -0.243 779\\ \hline

& 20A&  0.541 8660&  -0.306 983&  0.998&  0.242 910&  -0.240 163\\
& 20B&  0.541 9474&  -0.306 949&  0.991&  0.243 008&  -0.240 741\\ \hline

& 22A&  0.535 3653&  -0.304 979&  0.994&  0.240 278&  -0.237 650\\ \hline

& 24A&  0.529 5492&  -0.303 416&  0.998&  0.237 729&  -0.234 435\\
& 24B&  0.529 8510&  -0.303 301&  0.986&  0.238 082&  -0.235 625\\
& 24C&  0.529 3639&  -0.303 490&  0.981&  0.237 520&  -0.237 700\\
& 24D&  0.529 9500&  -0.303 263&  0.971&  0.238 197&  -0.236 017\\
& 24E&  0.529 9674&  -0.303 259&  0.969&  0.238 217&  -0.236 080\\
& 24F&  0.528 5647&  -0.303 799&  0.706&  0.236 597&  -0.230 544\\ \hline

& 26A&  0.525 1659&  -0.301 860&  0.964&  0.236 282&  -0.234 103\\
& 26B&  0.524 2474&  -0.302 191&  0.931&  0.235 239&  -0.231 066\\ \hline

& 28A&  0.520 7211&  -0.300 755&  0.993&  0.234 316&  -0.231 678\\
& 28B&  0.520 0785&  -0.300 975&  0.952&  0.233 596&  -0.229 824\\
& 28C&  0.521 1685&  -0.300 593&  0.910&  0.234 819&  -0.232 982\\
& 28D&  0.521 2568&  -0.300 557&  0.879&  0.234 179&  -0.233 244\\
& 28E&  0.519 3081&  -0.301 240&  0.688&  0.232 733&  -0.227 603\\
& 28F&  0.518 9486&  -0.301 366&  0.508&  0.232 332&  -0.226 563\\ \hline

& 30A&  0.516 8641&  -0.299 770&  0.999&  0.232 659&  -0.229 789\\
& 30B&  0.516 9294&  -0.299 758&  0.998&  0.232 731&  -0.230 024\\
& 30C&  0.516 9302&  -0.299 758&  0.997&  0.232 732&  -0.230 028\\
& 30D&  0.516 3136&  -0.299 952&  0.954&  0.232 050&  -0.228 439\\
& 30E&  0.517 3857&  -0.299 593&  0.925&  0.233 237&  -0.231 225\\
& 30F&  0.517 7871&  -0.299 441&  0.779&  0.233 683&  -0.232 289\\
& 30G&  0.514 6583&  -0.300 501&  0.281&  0.230 221&  -0.224 155\\ \hline

& 32A&  0.513 6234&  -0.298 855&  1.000&  0.231 396&  -0.228 763\\
& 32B&  0.513 4834&  -0.298 917&  0.999&  0.231 243&  -0.228 414\\
& 32C&  0.513 4433&  -0.298 928&  0.999&  0.232 340&  -0.228 320\\
& 32D&  0.513 4277&  -0.298 930&  0.995&  0.232 325&  -0.228 286\\
& 32E&  0.513 7568&  -0.298 812&  0.982&  0.231 542&  -0.229 141\\
& 32F&  0.513 8191&  -0.298 780&  0.969&  0.231 611&  -0.229 962\\
& 32G&  0.514 1207&  -0.298 684&  0.920&  0.231 941&  -0.229 957\\
& 32H&  0.514 1236&  -0.298 684&  0.908&  0.231 945&  -0.229 962\\
& 32I&  0.512 8576&  -0.299 116&  0.908&  0.231 739&  -0.226 930\\
& 32J&  0.514 3656&  -0.298 600&  0.829&  0.232 778&  -0.226 091\\
& 32K&  0.514 5824&  -0.298 521&  0.786&  0.231 709&  -0.229 962\\
& 32L&  0.512 5163&  -0.299 255&  0.737&  0.230 104&  -0.219 460\\
& 32M&  0.511 6581&  -0.299 480&  0.478&  0.229 246&  -0.224 108\\
& 32N&  0.511 5721&  -0.299 515&  0.426&  0.230 421&  -0.223 898\\
& 32P&  0.510 2774&  -0.299 915&  0.002&  0.229 101&  -0.220 851\\
& 32Q&  0.509 6882&  -0.300 104&  0.000&  0.233 290&  -0.230 544\\ \hline
\end{tabular} \end{center} }
\end{table}

A prime example of the importance of topology rather than geometry
is found in N = 32 bipartite bcc lattice 32J.  Each vertex has
eight first neighbours, fourteen topologically second neighbours,
eight topologically third neighbours and one topologically fourth
neighbour.  For all fourteen second neighbour pairs, 
$<{\bf S}_o\cdot {\bf S}_i>$
= 0.233290, although {\it geometrically} only six of these fourteen
neighbours are second neighbours, six are third neighbours and
two neighbours are fifth neighbours to the vertex chosen as origin.
 All topologically third neighbour pairs are geometrically fourth
neighbour pairs and all have the same spin-spin correlation,
similarly all first neighbour pairs have the same correlation,
as in Table 4.  As a geometric entity this lattice has rotationally
complete cubic or octahedral symmetry, $O_h$.  However, as a topological
entity lattice 32J has a still {\it greater} symmetry demonstrated
by the second neighbour correlations.

To obtain estimates of a physical property of the S = 1/2 Heisenberg
antiferromagnet on the infinite bcc lattice at zero temperature
we first fit a formula in inverse powers of L ($L^3 = N$) to the
ground state data for that property on each of the topologically
distinct finite bcc lattices of $N \le 32$ vertices.  
For instance, spin wave theory [1]
and other studies [18,19,20] show that the dimensionless ground
state energy per vertex, $\epsilon_0 \equiv E_0/NJ$, fits the formula

\[  \epsilon_0(L) =  \epsilon_0(\infty) + A_4 L^{-4} + A_6 L^{-6} + . . .
  \hspace{3cm}               (4.2) \]
Because $\epsilon_0$ is simply four times the average of the nearest neighbour
correlations, we use the same formula to fit the topologically
second and third neighbour correlations.

The dimensionless staggered magnetization operator,

\[ {\bf M}^+ \equiv \sum_{i=1}^{N/2}{\bf S}_i - \sum_{j=1}^{N/2}{\bf S_j}
  \hspace{4cm}  (4.3) \]
where the ${\bf S}_i$ are on one sublattice and the ${\bf S}_j$ on the other.
In the absence of an external field $< {\bf M}^+ > = 0$, but 

\[
 \left\langle({\bf M}^+)^2\right\rangle =
   \sum_{i,j=1}^{N}|\langle{\bf S}_i\cdot{\bf S}_j\rangle|
  \hspace{4cm}         (4.4) \]
is nonzero.  The staggered magnetization per vertex , $m^+$, is
calculated using

\[ m^+ = \left[\left\langle({\bf M}^+)^2\right\rangle\right]^{1/2}/N
  \hspace{4cm}        (4.5)\]
Spin wave theory[1] shows only that

\[ m^+(L)=m^+(\infty)+B_2L^{-2}+... \hspace{4cm}
                                  (4.6) \]
After some testing of various powers of L in a statistical analysis
of the data, we have settled on using as a third term $B_4L^{-4}$;
we have also tested a two parameter fit.  

Our fitting was done using the statistical programming package
S-PLUS (produced by MathSoft Inc., Seattle, USA).  We have also
obtained valuable advice from Wade Blanchard, an expert statistical
analyst in Dalhousie's Department of Mathematics and Statistics.
 First we perform for each property a standard least squares
fit of the data from all the topologically distinct lattices
to the appropriate formula.  Then each data point is assigned
a weight, $\sin(u)/u$, determined by the Huber weight function [21].
 The weights range from 1 for a point directly on the best fitted
curve to 0 for a distant outrider.  Weights for the energies
are shown in Table 4.

Blanchard advised us that cutoff weights are usually about 0.80,
although this cutoff depends on the data being used.  We have
varied the energy cutoff weights, \(w_c\), from as high as 0.95, which
would classify 19 of the 42 distinct lattices as outriders, to
as low as 0.75, which would classify only 9 of the lattices as
outriders.  The estimates of both the energy and staggered magnetization
peak as a function of \(w_c\) when \(w_c\) = 0.85 or, alternatively, where
the number of outriders, $N_o$, is 12.  Although the energy and
staggered magnetization weights are similar, we decided to use
the energy weights to define outriders because the energy is
simply the ground state eigenvalue of the Hamiltonian matrix,
where the staggered magnetization is calculated from the ground
state eigenvector.  Thus for each N we were able to rank those
bcc lattices that were not outriders with bcc lattice NA being
``best'', lattice NB second best, etc., as seen in Table 2. The
confidence limits that we have inserted in Table 5 after the
estimates of most properties are determined as the difference
between the estimates for $N_o$ = 12 and the estimates for $N_o$ =
11 and $N_o$ = 13.

\begin{table}
Table 5. Estimates (with confidence limits) of the T = 0 properties
of the S = 1/2 Heisenberg antiferromagnet on the infinite bcc
lattice
\begin{center}{ \small
\begin{tabular}{lllllllll} \hline\hline
Method  & $-\epsilon_0$  & $-A_4$  & $A_6$  &  $m^+$  & $B_2$   &  $B_4$  & $\Gamma_2$   &  $-\Gamma_3$ \\ \hline
finite lattice  & 1.1518(9)  & 5.24(10)  & 8.2(6)  & 0.4409(11) & 0.70(1)  & 0.30(7)  & 0.2161(6)  & 0.214(3) \\
spin wave [1]   & 1.1512(1)  & 4.5       &   ---   & 0.4412(3)  & 0.72     &    ---   &     ---    &       ---\\
series [1]      & 1.1510(5)  &   ---     &   ---   & 0.442(4)   &  ---     &    ---   &     ---    &       ---\\
variation [11]  & 1.160      &   ---     &   ---   & 0.426      &  ---     &    ---   &     ---    &       ---\\ \hline
\end{tabular} }\end{center}
\end{table}

Table 5 displays our finite lattice method estimates at zero
temperature of the physical properties of the spin one-half Heisenberg
antiferromagnet on the infinite body centred cubic lattice together
with estimates by three other methods.  According to the variational
estimate of the energy per vertex, $\epsilon_o$, the other three estimates
are too high by at least one percent, although they agree with
one another to within 0.05 \%.  The finite lattice estimate of
the staggered magnetization per vertex, $m^+$, agrees with the third
order spin wave and series expansion estimates to within 0.25
\%.  Some readers may notice that the spin wave estimates of
$A_4$ and $B_2$ displayed in this table are different from those in
[1].  The reason is that the authors in that article define $L$
as $L^3 = N/2$ .

The statistical analysis of the second and third neighbour correlations
on the infinite lattice have been made not only directly but
also by analysis of the ratios of, and differences between, the
first, second, and third neighbour correlations.  The resulting
variation among the estimates of 2 and 3 led to our confidence
limits.  To the best of our knowledge, no other estimates of
these correlations have been published.

Following the example of Oitmaa et al [1] we can use our estimates
in Table 5 to calculate other properties.  The spin wave velocity
$v = -A_4 / \beta$ where, using our definition of $L$ and the geometric
quantity of Hasenfratz and Leutwyler [19], $\beta$ = 2.1104607 so our
estimate is $v$ = 2.48(15).  Another geometric property of the
bcc lattice [19] is $\gamma$ = 0.17920577.  Then our estimate of the
spin stiffness, $\rho_s = m^+(\infty) v \gamma / B_2 = 0.280(11)$.  
Finally, the perpendicular
susceptibility $\chi_\perp = \rho_s / v^2 = 0.046(2)$.  
The most direct estimate
among the above three, for us and for Oitmaa et al [1], is that
of the spin wave velocity, $v$.  The third order spin wave estimate
2.2 is as good as we might expect in view of the large confidence
limits in each case.

\subsubsection*{5.  Summary, Conclusions and Outlook}

Following the example of earlier definitions, first of finite
simple cubic lattices [8] and second of finite face centred cubic
lattices [9], we have in this paper defined finite bipartite
body centred cubic lattices.  In each step improvements have
been made, first in introducing finite lattices in three dimensions,
next on introducing the defining vectors in the upper triangular
lattice form and now in classifying bipartite finite lattices
{\it topologically}, an important step beyond geometric classification.
 Indeed to establish our topological classification we have introduced
an entity, the {\it topological neighbourhood matrix}.

Geometrically distinct but topologically identical finite lattices
have neighbourhood matrices that look superficially distinct
but are mathematically identical.  Because for finite lattices
with a small number of vertices the neighbourhood matrices are
very simple, we have been able to derive from them a simple topological
code to sort out finite bipartite bcc lattices, a type of sorting
code that would work equally well on simple cubic or other lattices.

We have also used the old but largely unfamiliar Smith normal
form of the neighbourhood matrix as an alternate way to sort
finite lattices topologically.  The Smith matrix is much more
complicated to derive from the neighbourhood matrix than is our
sorting code, but it would work for quite large lattices well
beyond the scope of our simpler method.  When both methods answered
that two lattices were topologically distinct, the ground state
properties of the Heisenberg Hamiltonian on these two lattices
invariably confirmed this fact.

We have diagonalized the spin one-half Heisenberg antiferromagnet
Hamiltonian on all geometrically distinct bipartite bcc lattices
of sixteen to thirty vertices.  Using both our methods of recognizing
topologically identical bipartite lattices, we have diagonalized
the Hamiltonian on only sixteen topologically distinct thirty-two
vertex bipartite bcc lattices, thus saving many hours of computing
time.  The high performance computer used standard procedures
to compute to very high precision the ground state eigenvalue
(energy) and eigenvector on each lattice.  Thence all spin-spin
correlations and the staggered magnetization were derived, and
other properties such as four-spin correlations could have been
derived.  

These data for each physical property were fitted statistically
to appropriate equations using inverse powers of L, the cube
root of the number of vertices.  Unlike some methods, our finite
lattice method enables the determination (statistically) of the
confidence limits of the estimates of each property calculated.

We were pleased to find that our estimate of the energy per vertex
of the Heisenberg antiferromagnet on the infinite bcc lattice
at zero temperature agrees with the third order spin wave and
series expansion estimates of Oitmaa et al [1] to within five
parts in ten thousand.  Our estimate of the staggered magnetization
agrees with the spin wave estimate [1] to within a quarter of
one percent and within the one percent confidence limits of the
series estimate.  Variational estimates of energy and staggered
magnetization [11] differ from those of the other three methods
by a larger amount.  We have not found calculations by other
methods of the second and third neighbour spin-spin correlations
that we have calculated, but our estimates are useful because
the correlations give insight into the nature of the ground state
eigenvector.

We would like to see by other theoretical methods, such as the
quantum Monte Carlo method,  estimates of the properties we have
calculated.  Also we failed to find in the literature experimental
measurements of energy and staggered magnetization at near zero
temperature on magnetic materials that can be well represented
by the spin one half Heisenberg antiferromagnet on the body centred
cubic lattice.  

A nearly ideal three dimensional isotropic Heisenberg antiferromagnet
is the magnetic system of RbMnF$_3$ [22], but the magnetic moments
have 5/2 spins.  Experimental examples of a spin one-half isotropic
Heisenberg antiferromagnet on a bcc lattice have been hard to
find, but very recently Srdanov et al [23] have found evidence
for a spin one-half Heisenberg bcc antiferromagnet consisting
of F centres in sodium-electro-sodalite.

Perhaps the greatest advance described in this paper is the recognition
of the importance of topology in the theoretical study of quantum
spin systems at zero temperature.  We have learned much else
in the past two or three years, but there is much more to learn
and do in this corner of physics -- studying other lattices,
different properties, higher spin, non-zero temperature, etc.
 More powerful computers would help us, and they will become
available.  Better still, we invite theoretical and experimental
colleagues to join us in our exploration.

Acknowledgement.  We are grateful to colleagues H. Nishimori,
P. Keast, W. Blanchard and W. Zheng for their useful advice.
 Part of this paper was written when one of the authors (D.D.B)
was a Visiting Professor at the University of New South Wales.
 This research has been supported in part by the Natural Sciences
and Engineering Research Council of Canada and by Deutsche Forschungsgemeinschaft
(project Ri 615/1-2)

\subsubsection*{References}
\begin{description}
\item[{[1]}]  Oitmaa J, Hamer C J and Zheng W 1994 {\it Phys. Rev.} B {\bf 50} 3877
\item[{[2]}]  Bonner J and Fisher M E 1964 {\it Phys. Rev.} {\bf 153} A640
\item[{[3]}]  Betts D D and Oitmaa J 1977 {\it Phys. Lett. A} {\bf 62} 277
\item[{[4]}]  Oitmaa J and Betts D D 1978 {\it Can. J. Phys.} {\bf 56} 897
\item[{[5]}]  Manousakis E 1991 {\it Rev. Mod. Phys.} {\bf 63} 1
\item[{[6]}]  Dagatto E 1994 {\it Rev. Mod. Phys.} {\bf 66} 763
\item[{[7]}]  Oitmaa J and Betts D D 1978 {\it Phys. Lett. A} {\bf 68} 450
\item[{[8]}]  Betts D D and Stewart G E 1977 {\it Can. J. Phys.} {\bf 75} 47
\item[{[9]}]  Stewart G E, Betts D D and Flynn J S 1997 {\it J. Phys. Soc. Japan}
	{\bf 66} 3231
\item[{[10]}]  Lyness J N, Sorevik T and Keast P 1991 {\it Mathematics of
	Computation} {\bf 56} 243
\item[{[11]}]  Ol\`{e}s A M and Ol\`{e}s B 1993 {\it J. Phys. : Condens. Matter}
	{\bf 5} 8403
\item[{[12]}]  Smith H J S 1861 {\it Phil. Trans.} {\bf 151} 293
\item[{[13]}]  Turnbull H W and Aitken A C 1932 {\it An Introduction to the
	Theory of Canonical Matrices} (Glasgow: Blackie) 21-23
\item[{[14]}]  Lieb E H and Mattis D C 1962 {\it J. Math. Phys.} {\bf 3} 749
\item[{[15]}]  Kennedy T, Lieb E H and Shastry B S 1988 {\it J. Stat. Phys.}
	{\bf 53} 1019
\item[{[16]}]  Kubo K and Kishi T 1988 {\it Phys. Rev. Lett.} {\bf 61} 2585
\item[{[17]}]  Cullum J K and Willoughby R A 1985 {\it Lanczos Algorithms
	for Large Symmetric Eigenvalue Computations } (Birkhhaeuser)
\item[{[18]}]  Neuberger H and Ziman T 1989 {\it Phys. Rev. B}  {\bf 39} 2608
\item[{[19]}]  Fisher D S 1989 {\it Phys. Rev. B}  {\bf 39} 11783
\item[{[20]}]  Hasenfratz P and Leutwyler H 1990 {\it Nuclear Phys. B}  {\bf 343} 241
\item[{[21]}]  Hoaglin D C, Mostelle F and Tukey J W 1983 {\it Understanding
	Robust and Exploratory Data Analysis} (New York: John Wiley \&
	Sons) 366
\item[{[22]}]  Coldea R, Cowley R, Ferring T G, McMorrow D F and Roessli
	B 1998 {\it Phys. Rev. B} {\bf 57} 5281
\item[{[23]}]  Srdanov V I, Stucky G D, Lippmaa E and Engelhardt G 1998
	{\it Phys. Rev. Lett.} {\bf 80} 2449
\end{description}

\subsubsection*{List of Figures}
Figure 1.  A sample eighteen-vertex, bcc lattice-filling parallelepiped
is defined in this Figure by the eight corner vertices and the
dashed lines between them representing the edges.  Application
of periodic boundary conditions to this parallelepiped forms
the bipartite N = 18 bcc lattice.  The nine vertices on one sublattice,
labelled by capital letters, are represented by black circles,
and the vertices on the other sublattice, labelled by lower case
letters, are represented by white squares.  The solid lines connect
vertex H to its eight nearest neighbours.
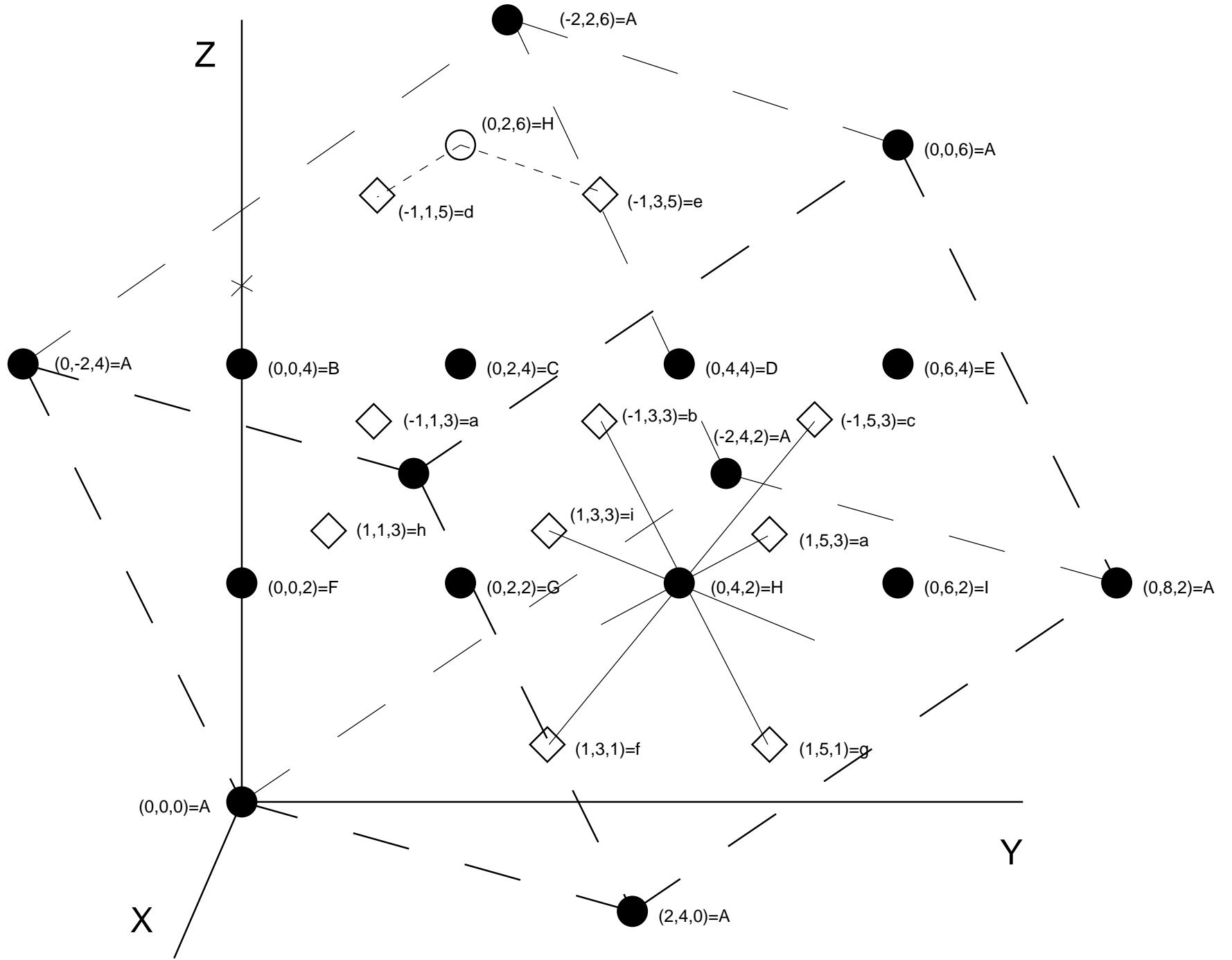
\end{document}